\documentclass[11pt]{article}
\usepackage[a4paper,margin=0.88in]{geometry}
\usepackage{amsmath,amssymb,microtype}
\usepackage{graphicx}
\usepackage[hidelinks]{hyperref}
\usepackage{enumitem}
\title{Deriving the Kijowski Arrival-Time POVM from the Schr\"odinger Current: Minimal Positivity and Uniqueness}
\author{
Avi Marchewka\\
\texttt{Avi.marchewka@gmail.com}
}

\begin{document}
\maketitle

\begin{abstract}
Quantum backflow refers here to the appearance of a negative Schr\"odinger
current for a state whose momentum support is entirely positive. We ask for the smallest modification of the free Schr\"odinger current that
makes it nonnegative for every such state, while preserving the current of
each individual momentum component. We show that the required minimal modification changes the free-particle
momentum kernel according to
\[
K_{\rm Sch}(p,p')=\frac{p+p'}{2m}
\;\longrightarrow\;
K_{\min}(p,p')=\frac{\sqrt{pp'}}{m}.
\]
The resulting current is positive and normalized and therefore defines an
arrival-time POVM. Extending the directional no-backflow requirement to
states containing both momentum signs forces the cross-sector kernel to
vanish,
\[
K_{\min}^{+-}=K_{\min}^{-+}=0,
\]
so that the full current is the sum of two independent directional
contributions. The resulting POVM is exactly the Kijowski time-of-arrival
POVM, providing a current-based physical motivation for both its
directional kernels and their separation. Within the diagonal-preserving
pairwise-minimal construction considered here, the result is unique. The
construction itself does not impose a first-arrival condition.
\end{abstract}

 \section{Introduction}
 The Schrödinger current follows directly from the continuity equation and is the natural local probability current of a freely evolving particle. However, quantum backflow shows that even for states whose momentum support lies entirely in \(p>0\), the current can become negative \cite{BrackenMelloy1994,MugaLeavens2000}. It therefore cannot, in general, serve directly as a positive arrival-time probability density.

A different approach to the arrival-time problem for a free particle was introduced by Kijowski. Kijowski constructs an arrival-time distribution axiomatically \cite{Kijowski1974,Werner1987,Kijowski1999}. In one dimension, the resulting density has the dimensions of probability per unit time, as does the probability current through a point, and separates into two directional contributions associated with positive and negative momenta. Together these form the normalized Kijowski POVM. The construction addresses arrival time, but does not impose a first-arrival condition.

In the present work we return directly to the current and ask a simple
question: how should the free Schr\"odinger-current kernel be modified so
that the resulting current is nonnegative, and can such a modification
define an arrival-time POVM? We show that the answer to both questions is
positive. On each fixed-sign momentum sector, the minimal modification
required to remove backflow yields a positive normalized current. Extending
the same directional no-backflow requirement to states containing both
momentum signs forces the cross-sector kernel to vanish, so that the full
current is the sum of two independent directional contributions. The
resulting POVM is exactly the Kijowski arrival-time POVM.

The structure of the paper is as follows. Section 2 introduces the
Schr\"odinger current and its momentum-space kernel. Section 3 derives the
minimal positive kernel on a fixed-sign momentum sector and discusses its
principal properties. Section 4 extends the construction to the opposite
and mixed momentum sectors and shows that the cross-sector kernel must
vanish. Section 5 identifies the resulting full positive current with the
Kijowski arrival-time POVM. Section 6 summarizes the physical implications,
including the relation to constrained and first-arrival constructions.

\section{The fixed-sign Schr\"odinger current}
For a free particle in one dimension,
\begin{equation}
 i\hbar\partial_t\psi(x,t)=-\frac{\hbar^2}{2m}\partial_x^2\psi(x,t),
\end{equation}
and the density $\rho=|\psi|^2$ satisfies
\begin{equation}
 \partial_t\rho+\partial_x j_{\rm Sch}=0,
 \qquad
 j_{\rm Sch}(x,t)=\frac{\hbar}{m}\operatorname{Im}\!\left[\psi^*(x,t)\partial_x\psi(x,t)\right].
\end{equation}
We choose the observation point $x=0$.  Before restricting the momentum sign, it is useful to make explicit how the two directional sectors arise from a completely general free state,
\begin{equation}
 \psi(x,t)=\frac{1}{\sqrt{2\pi\hbar}}
 \int_{-\infty}^{\infty}g(p)
 e^{ipx/\hbar-iE_pt/\hbar}\,dp,
 \qquad E_p=\frac{p^2}{2m}.
\end{equation}
For $p>0$ define the even and odd combinations
\begin{equation}
 g_e(p)=g(p)+g(-p),\qquad
 g_o(p)=g(p)-g(-p).
\end{equation}
The transformation is invertible,
\begin{equation}
 g(p)=\frac{g_e(p)+g_o(p)}{2},\qquad
 g(-p)=\frac{g_e(p)-g_o(p)}{2},
\end{equation}
so no information about the two momentum directions has been discarded.  Writing $\psi=\psi_e+\psi_o$,
\begin{align}
 \psi_e(x,t)&=\frac{1}{\sqrt{2\pi\hbar}}
 \int_0^\infty g_e(p)\cos(px/\hbar)e^{-iE_pt/\hbar}\,dp,\\
 \psi_o(x,t)&=\frac{i}{\sqrt{2\pi\hbar}}
 \int_0^\infty g_o(p)\sin(px/\hbar)e^{-iE_pt/\hbar}\,dp.
\end{align}
At the origin, $\psi_o(0,t)=0$ and $\partial_x\psi_e(0,t)=0$, so the current is carried entirely by even--odd interference,
\begin{equation}
 j_{\rm Sch}(0,t)=\frac{\hbar}{m}
 \operatorname{Im}\!\left[\psi_e^*(0,t)\,\partial_x\psi_o(0,t)\right].
\end{equation}
The two fixed-sign sectors correspond simply to
\begin{equation}
 g(-p)=0\iff g_e(p)=g_o(p),\qquad
 g(p)=0\iff g_e(p)=-g_o(p),\qquad p>0.
\end{equation}
Thus the use of $p>0$ below is not a loss of the opposite direction; it is the choice of one of the two directional backflow problems.  In the positive-momentum sector set $g_e=g_o=f$.  Then the current takes the compact form
\begin{equation}
 j_{\rm Sch}^{(+)}(t)=\frac{1}{2\pi\hbar}
 \int_0^\infty\!\!\int_0^\infty
 f^*(p)\,K_{\rm Sch}(p,p')\,f(p')
 e^{i(E_p-E_{p'})t/\hbar}\,dp\,dp'.
\end{equation}
with
\begin{equation}
 \boxed{K_{\rm Sch}(p,p')=\frac{p+p'}{2m}.}
\end{equation}
The diagonal is
\begin{equation}
 K_{\rm Sch}(p,p)=\frac{p}{m}>0.
\end{equation}
Thus every individual momentum component has the correct positive flux.  Only the off-diagonal terms can reverse the total current.

\section{Minimal positivity with the diagonal fixed}
 Consider first two positive momenta $p_1,p_2$. Keeping their individual
fluxes fixed, a real symmetric corrected kernel\footnote{
For simplicity we write the corrected kernel as real symmetric, as is the
free Schr\"odinger-current kernel being modified. Allowing the more general
Hermitian form with off-diagonal elements $B$ and $B^*$ does not change the
result: positivity gives $|B|\le \sqrt{p_1p_2}/m$, and the point in this disk
closest to the real Schr\"odinger value $(p_1+p_2)/(2m)$ is uniquely
$B=\sqrt{p_1p_2}/m$.
}
has the form   
\begin{equation}
 K^{(2)}=
 \begin{pmatrix}
 p_1/m & B\\
 B & p_2/m
 \end{pmatrix}.
\end{equation}
Positivity for every two-component superposition requires
\begin{equation}
 \det K^{(2)}=\frac{p_1p_2}{m^2}-B^2\ge0,
\end{equation}
so
\begin{equation}
 |B|\le\frac{\sqrt{p_1p_2}}{m}.
\end{equation}
The Schr\"odinger coefficient is
\begin{equation}
 B_{\rm Sch}=\frac{p_1+p_2}{2m}\ge\frac{\sqrt{p_1p_2}}{m},
\end{equation}
with equality only for $p_1=p_2$.  For the fixed diagonal, positivity therefore restricts the real off-diagonal coefficient to the interval
\begin{equation}
 -\frac{\sqrt{p_1p_2}}{m}\le B\le\frac{\sqrt{p_1p_2}}{m}.
\end{equation}
The Schr\"odinger value lies at or to the right of this entire admissible interval.  Hence the admissible coefficient requiring the smallest change from the Schr\"odinger value is uniquely the upper endpoint,
\begin{equation}
 \boxed{B_{\min}=\frac{\sqrt{p_1p_2}}{m}.}
\end{equation}
Equivalently, the required pairwise correction is
\begin{equation}
 B_{\rm Sch}-B_{\min}
 =\frac{(\sqrt{p_1}-\sqrt{p_2})^2}{2m}\ge0.
\end{equation}

This is the precise two-momentum sense in which the correction is minimal:
the one-momentum fluxes are held fixed, and the coherence coefficient is
moved only as far as positivity requires.

Since every positive kernel with the prescribed diagonal satisfies
\[
|K(p,p')|\leq \frac{\sqrt{pp'}}{m},
\]
and the Schr\"odinger coefficient lies above this bound, the
geometric-mean kernel is pointwise the closest admissible choice for every
momentum pair. Since these pairwise choices together form a positive
continuum kernel, the minimization is simultaneously realizable for all
pairs.

Since the two-momentum result holds for an arbitrary pair $p,p'>0$, we use
the same pairwise-minimal coefficient throughout the positive-momentum
sector,
\begin{equation}
\boxed{K_{\min}(p,p')=\frac{\sqrt{pp'}}{m}.}
\end{equation}

Substituting this kernel directly into the continuum current gives
\begin{align}
 J_{\min}(t)
 &=\frac{1}{2\pi\hbar m}
 \int_0^\infty\!\!\int_0^\infty
 \sqrt p\,\sqrt{p'}\,f^*(p)f(p')
 e^{i(E_p-E_{p'})t/\hbar}\,dp\,dp'\\
 &=\frac{1}{2\pi\hbar m}
 \left|\int_0^\infty\sqrt p\,f(p)e^{-iE_pt/\hbar}\,dp\right|^2\ge0.
\end{align}
Thus the pairwise-minimal coefficients are compatible with the full continuum: taken together, they produce a single positive current for every state in the sector.

\medskip
\noindent\textbf{Emergent POVM structure.}
The positivity requirement was imposed on the Schr\"odinger current while its diagonal was kept fixed.  Since the full-time integral of the free current selects precisely this diagonal, a normalized state in the positive-momentum sector satisfies
\begin{equation}
 \int_{-\infty}^{\infty}J_{\min}(t)\,dt
 =\int_{-\infty}^{\infty}j_{\rm Sch}^{(+)}(t)\,dt
 =\int_0^\infty |f(p)|^2\,dp=1.
\end{equation}
Hence the minimal modification does more than remove negative values: it converts the normalized but sign-indefinite Schr\"odinger current into a positive normalized operator-valued density.  The POVM structure therefore emerges from minimal positivity together with preservation of the Schr\"odinger diagonal; its identification with the Kijowski POVM is made in Sec.~5.

\medskip
  \noindent\textbf{Coherence attenuation.}
The modification acts only on unequal-momentum coherence. Relative to the
Schr\"odinger kernel,
\begin{equation}
 K_{\min}(p,p')=C(p,p')K_{\rm Sch}(p,p'),
 \qquad
 \boxed{C(p,p')=\frac{2\sqrt{pp'}}{p+p'}\leq 1.}
\end{equation}
On the diagonal $C(p,p)=1$, whereas for $p\neq p'$ one has
$0<C(p,p')<1$. For a fixed momentum pair,
\begin{equation}
 j_{pp'}^{\min}(t)
 =
 C(p,p')\,j_{pp'}^{\rm Sch}(t).
\end{equation}
Thus the same attenuation factor reduces a positive coherent contribution
while making a negative one less negative; in both cases it reduces the
magnitude of the coherent contribution toward zero. By the result above,
this is precisely the minimal attenuation required to make the current
nonnegative, thereby removing backflow.

\begin{figure}[t]
 \centering
 \includegraphics[width=0.58\linewidth]{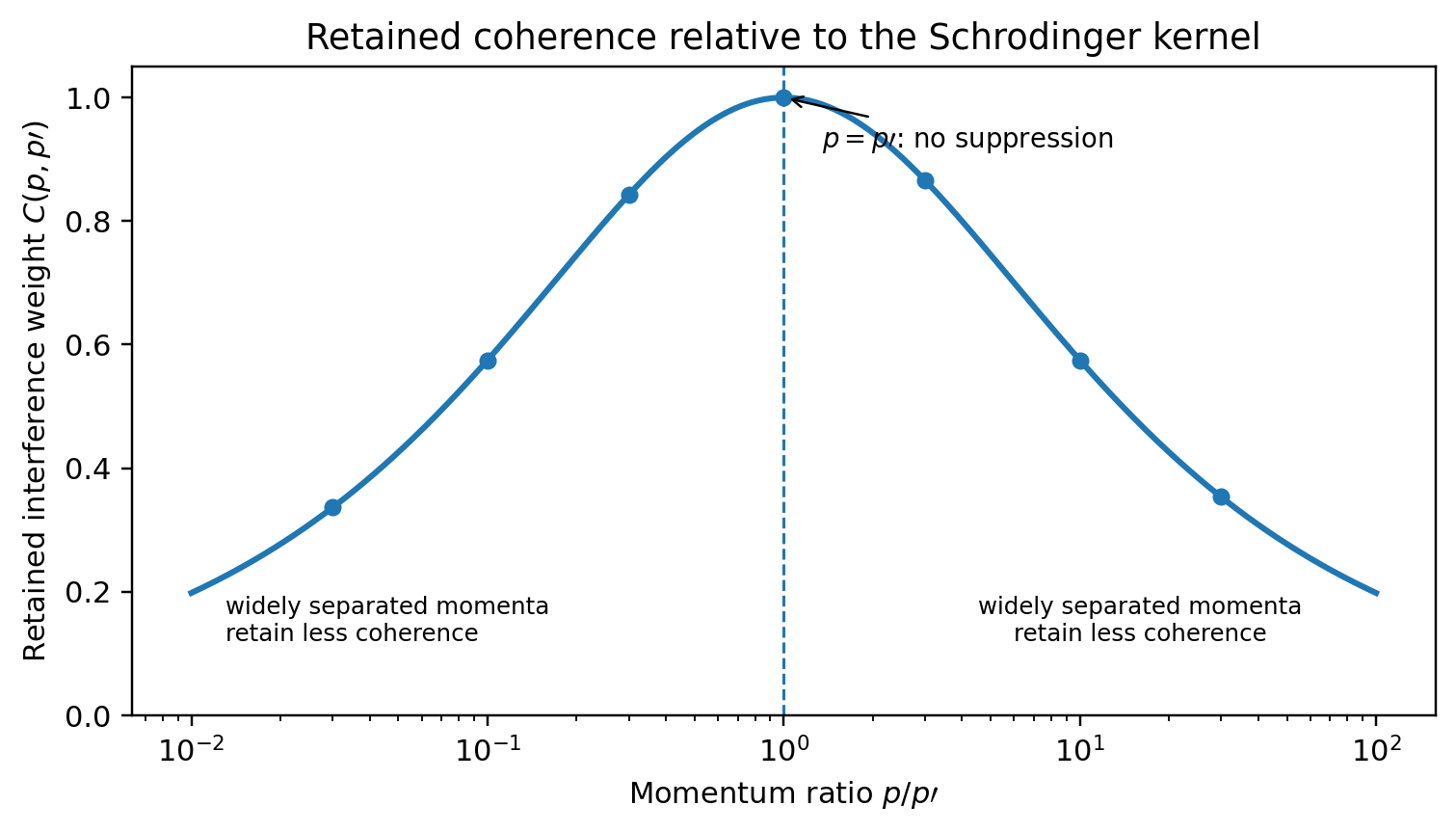}
 \caption{The coherence attenuation factor
 $C(r)=2\sqrt r/(1+r)$. It is unity for equal momenta and decreases
 symmetrically as the ratio of the two momenta departs from unity.}
 \label{fig:coherence}
\end{figure}

The dependence of the attenuation on the separation of the two momenta is
seen more clearly by writing
\begin{equation}
 r=\frac{p}{p'},
 \qquad
 C(r)=\frac{2\sqrt r}{1+r}.
\end{equation}
It satisfies
\begin{equation}
 C(1)=1,
 \qquad
 C(r)=C(r^{-1}),
 \qquad
 C(r)\rightarrow0
 \quad\text{as}\quad r\rightarrow0\ \text{or}\ \infty .
\end{equation}
Thus nearby momentum components are only weakly attenuated, whereas
coherence between widely separated momenta is strongly suppressed. These properties are illustrated in Fig.~\ref{fig:coherence}: the attenuation is weak for nearby momenta and becomes progressively stronger as their ratio departs from unity.

\medskip
\noindent\textbf{Difference between the kernels and redistribution in time.}
The pointwise difference is
\begin{equation}
 \Delta K(p,p'):=K_{\rm Sch}(p,p')-K_{\min}(p,p')
 =\frac{(\sqrt p-\sqrt{p'})^2}{2m}\ge0,
\end{equation}
with $\Delta K(p,p)=0$.  Since $p,p'>0$, both kernels are pointwise nonnegative and $K_{\min}\le K_{\rm Sch}$.  This pointwise ordering does \emph{not} imply an ordering of the currents.  The kernel $\Delta K$ is not positive semidefinite as an integral kernel, so the corresponding coherent contribution can have either sign.  Consequently $J_{\min}(t)$ can be either smaller or larger than $j_{\rm Sch}^{(+)}(t)$ at a given time, while
\begin{equation}
 \boxed{\int_{-\infty}^{\infty}
 \left[j_{\rm Sch}^{(+)}(t)-J_{\min}(t)\right]dt=0.}
\end{equation}
The minimal correction therefore redistributes the current in time rather than removing full-time weight.

 \medskip

\noindent\textbf{No first-arrival condition.}
The minimal-current construction defines a positive normalized arrival-time
density, but it does not impose a first-arrival condition. A first-arrival
density at time $t$ must refer to arrival at $t$ under the condition that no
earlier arrival has occurred. No such conditioning is present here: the
density is constructed directly from the freely evolved wave function.
In particular, nothing in the construction excludes a nonzero probability
density at the arrival point at an earlier time,
\begin{equation}
 \rho(x_D,s)=|\psi(x_D,s)|^2\neq 0,
 \qquad s<t.
\end{equation}
Thus the minimal current provides an arrival-time density, but first arrival
is not encoded in the construction itself.

\subsection{Two-mode illustration}

The effect of the minimal kernel can be seen explicitly in the simplest
two-momentum example.  Consider two positive momenta
\[
p_1=1,\qquad p_2=9,
\]
for which
\[
K_{\rm Sch}
=
\frac{1}{m}
\begin{pmatrix}
1 & 5\\
5 & 9
\end{pmatrix},
\qquad
K_{\min}
=
\frac{1}{m}
\begin{pmatrix}
1 & 3\\
3 & 9
\end{pmatrix}.
\]
Thus the diagonal terms are unchanged, while the coherent term is reduced by
\[
C(p_1,p_2)=\frac{3}{5}.
\]

Choose the normalized two-mode state
\[
|\psi_\theta\rangle
=
\frac{1}{\sqrt{10}}
\left(
3|p_1\rangle+e^{i\theta}|p_2\rangle
\right),
\]
where $\theta$ is the relative phase.  The two currents are then
\begin{equation}
j_{\rm Sch}(\theta)
=
\frac{1}{m}
\left(
\frac{9}{5}+3\cos\theta
\right),
\end{equation}
and
\begin{equation}
J_{\min}(\theta)
=
\frac{9}{5m}
\left(
1+\cos\theta
\right)
\ge 0.
\end{equation}

At maximal destructive interference, $\theta=\pi$,
\[
j_{\rm Sch}(\pi)=-\frac{6}{5m},
\qquad
J_{\min}(\pi)=0,
\]
so the negative Schr\"odinger current is removed.  At constructive
interference, $\theta=0$,
\[
j_{\rm Sch}(0)=\frac{24}{5m},
\qquad
J_{\min}(0)=\frac{18}{5m},
\]
showing that the same fixed attenuation also reduces a positive coherent
contribution.  The minimal modification therefore acts symmetrically on the
coherence amplitude: it reduces its magnitude sufficiently to eliminate
backflow while leaving the diagonal momentum contributions unchanged.

\begin{figure}[t]
 \centering
 \includegraphics[width=0.65\linewidth]{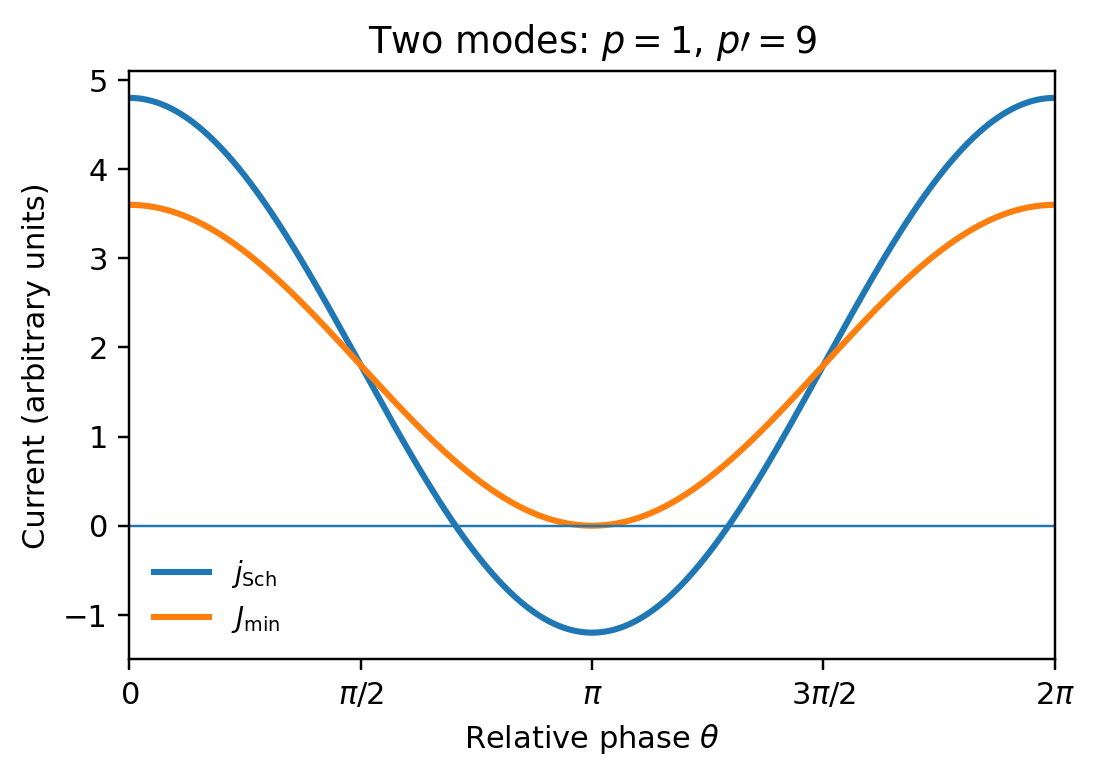}
 \caption{Two-mode illustration for $p_1=1$ and $p_2=9$.
 The minimal kernel reduces the coherent contribution by
 $C(p_1,p_2)=3/5$.  The Schr\"odinger current becomes negative near maximal
 destructive interference, whereas the minimal current remains nonnegative.
 The same attenuation also lowers the current in the constructive-interference
 region.}
 \label{fig:twomode}
\end{figure}

\section{Minimal-Current Construction for Opposite and Mixed Momentum Sectors}

The construction of Sec.~3 was carried out for states supported in the
positive-momentum sector. We now extend the same directional requirement
to negative momenta and then to states containing both momentum signs.
\medskip
\noindent

\textbf{Negative-momentum sector.}
For a state supported entirely in $p<0$, the direction of the Schr\"odinger
current is reversed. Writing
\[
p=-q,\qquad q>0,
\]
and defining the positive left-directed current by
\[
J_-(t)=-j_{\rm Sch}^{(-)}(t),
\]
the same argument as in Sec.~3 gives
\begin{equation}
 K_{\min}^{(-)}(q,q')
 =
 \frac{\sqrt{qq'}}{m}.
\end{equation}
Thus the minimal construction is identical in the two fixed-sign sectors
after reversing the orientation of the current.

\textbf{Opposite-sign momentum components.}
The same two-component argument can be applied to an arbitrary pair of
opposite momenta, $p>0$ and $-q<0$. For mixed-sign momentum states, the no-backflow condition is naturally
extended by requiring that the coherence between the two oppositely directed
momentum sectors not reverse the direction fixed by their diagonal
contributions. Then, on the two-dimensional subspace
spanned by $|p\rangle$ and $|-q\rangle$, a diagonal-preserving kernel has
the form
\begin{equation}
K^{(2)}(p,-q)
=
\begin{pmatrix}
p/m & B(p,q)\\
B^*(p,q) & -q/m
\end{pmatrix}.
\end{equation}

For the state
\begin{equation}
|\psi\rangle=c_+|p\rangle+c_-|-q\rangle,
\end{equation}
the corresponding current is
\begin{equation}
J
=
\frac{1}{m}\left(p|c_+|^2-q|c_-|^2\right)
+
2\,\mathrm{Re}\!\left[B(p,q)c_+^*c_-\right].
\end{equation}

Choose the amplitudes such that
\begin{equation}
p|c_+|^2=q|c_-|^2 .
\end{equation}
The diagonal contributions then cancel.  Since the relative phase between
$c_+$ and $c_-$ is arbitrary, any nonzero $B(p,q)$ would generate a current
of either sign entirely from the off-diagonal term.  The directional
no-backflow requirement therefore implies
\begin{equation}
\boxed{B(p,q)=0.}
\end{equation}

Thus, for every opposite-sign momentum pair, the directional-current kernel
has the form
\begin{equation}
K^{(2)}_{\rm dir}(p,-q)
=
\begin{pmatrix}
p/m & 0\\
0 & -q/m
\end{pmatrix}.
\end{equation}

Since $p>0$ and $q>0$ were arbitrary, the entire cross-sector kernel
vanishes,
\[
K^{+-}=K^{-+}=0.
\]
Consequently, there is no independent mixed-sector contribution to the
arrival current. A state containing both momentum signs is therefore
completely described by its two fixed-sign components, each retaining the
full within-sector kernel derived above. This result will provide the
current-based reason, below, for the separation of the two directional
branches in the Kijowski POVM.

\noindent

 \noindent\textbf{Two-component illustration.}
As a simple example, take
\begin{equation}
 |\psi_\theta\rangle
 =
 \frac{1}{\sqrt{10}}
 \left(|9\rangle+3e^{i\theta}|-1\rangle\right).
\end{equation}
The two diagonal directional contributions exactly cancel,
\begin{equation}
 J_0=\frac{1}{m}\left(9\frac{1}{10}
 -1\frac{9}{10}\right)=0,
\end{equation}
whereas the Schr\"odinger cross term gives
\begin{equation}
 j_{\rm Sch}(\theta)=\frac{12}{5m}\cos\theta .
\end{equation}
Thus the direction of the current is determined entirely by the relative
phase, even though the diagonal directional contributions are balanced.
The minimal mixed-sector prescription removes precisely this possibility,
since $K_{\min}(9,-1)=0$.
\medskip
\noindent

 \medskip
\noindent\textbf{Summary of the minimal-current construction.}
The minimal-current requirement therefore fixes both the kernel structure
and the corresponding directional currents.

For the positive-momentum sector,
\begin{equation}
K_{\min}^{++}(p,p')
=
\frac{\sqrt{pp'}}{m},
\end{equation}
and
\begin{equation}
J_+(t)
=
\frac{1}{2\pi\hbar m}
\left|
\int_0^\infty
\sqrt{p}\,f_+(p)
e^{-iE_pt/\hbar}\,dp
\right|^2
\ge 0 .
\end{equation}

For the negative-momentum sector, writing $q=-p>0$,
\begin{equation}
K_{\min}^{--}(q,q')
=
\frac{\sqrt{qq'}}{m},
\end{equation}
and the positive left-directed arrival current is
\begin{equation}
J_-(t)
=
\frac{1}{2\pi\hbar m}
\left|
\int_0^\infty
\sqrt{q}\,f_-(-q)
e^{-iE_qt/\hbar}\,dq
\right|^2
\ge 0 .
\end{equation}

 For opposite momentum signs, the directional no-backflow requirement gives
\begin{equation}
K_{\min}^{+-}=K_{\min}^{-+}=0,
\end{equation}
so that the two directional sectors contribute without interference. Any
mixed-sign momentum state is therefore completely represented by its
components in the two separate fixed-sign sectors.

The full positive arrival-time density generated by the minimal-current
construction is therefore
\begin{equation}
\boxed{
J_{\min}(t)=J_+(t)+J_-(t).
}
\end{equation}
Moreover,
\begin{equation}
\int_{-\infty}^{\infty}J_+(t)\,dt
=
\|P_+\psi\|^2,
\qquad
\int_{-\infty}^{\infty}J_-(t)\,dt
=
\|P_-\psi\|^2,
\end{equation}
and hence, for a normalized state,
\begin{equation}
\int_{-\infty}^{\infty}J_{\min}(t)\,dt=1.
\end{equation}
Thus the minimal-current requirement produces a positive normalized
arrival-time density consisting of two independent directional sectors. 
\medskip
\noindent

\section{Identification with the Kijowski POVM}
The minimal-current construction has fixed the full directional structure:
within each fixed-sign momentum sector the kernel is the geometric-mean
kernel, while the coherence between opposite momentum sectors vanishes.
We now compare this structure with the Kijowski arrival-time POVM.

For the positive-momentum sector, define
\begin{equation}
 |t,+\rangle=
 \frac{1}{\sqrt{2\pi\hbar m}}
 \int_0^\infty
 \sqrt p\,e^{iE_pt/\hbar}|p\rangle\,dp,
\end{equation}
and
\begin{equation}
 \Pi_+(t)=|t,+\rangle\langle t,+|.
\end{equation}
Its momentum-space kernel is
\begin{equation}
 \langle p|\Pi_+(t)|p'\rangle
 =
 \frac{\sqrt{pp'}}{2\pi\hbar m}
 e^{i(E_p-E_{p'})t/\hbar},
 \qquad p,p'>0,
\end{equation}
so that
\begin{equation}
 \langle\psi|\Pi_+(t)|\psi\rangle=J_+(t).
\end{equation}
Thus the positive-momentum minimal current obtained above is exactly the
positive-momentum Kijowski branch
\cite{Kijowski1974,Werner1987,Kijowski1999}.

For the negative-momentum sector, introducing $q=-p>0$,
\begin{equation}
 |t,-\rangle=
 \frac{1}{\sqrt{2\pi\hbar m}}
 \int_0^\infty
 \sqrt q\,e^{iE_qt/\hbar}|-q\rangle\,dq,
 \qquad
 \Pi_-(t)=|t,-\rangle\langle t,-|,
\end{equation}
and similarly
\begin{equation}
 \langle\psi|\Pi_-(t)|\psi\rangle=J_-(t).
\end{equation}
Hence the negative-momentum minimal current is exactly the negative-momentum
Kijowski branch.

The Kijowski POVM contains no cross terms between the two momentum-sign
sectors,
\begin{equation}
 \boxed{
 \Pi_K(t)=\Pi_+(t)+\Pi_-(t).
 }
\end{equation}
This is precisely the structure obtained above from the directional
no-backflow requirement,
\begin{equation}
 K_{\min}^{+-}=K_{\min}^{-+}=0.
\end{equation}
Thus, from the present current-based perspective, the separation of the two
Kijowski branches is consistent with the requirement that coherence between
oppositely directed momentum sectors must not generate a reversal of the
directional current.

Finally, using
\begin{equation}
 \int_{-\infty}^{\infty}\frac{dt}{2\pi\hbar}
 e^{i(E_p-E_{p'})t/\hbar}
 =
 \frac{m}{p}\delta(p-p'),
 \qquad p,p'>0,
\end{equation}
one obtains
\begin{equation}
 \int_{-\infty}^{\infty}\Pi_+(t)\,dt=P_+,
 \qquad
 \int_{-\infty}^{\infty}\Pi_-(t)\,dt=P_-,
\end{equation}
and therefore
\begin{equation}
 \boxed{
 \int_{-\infty}^{\infty}\Pi_K(t)\,dt
 =
 P_++P_-=I.
 }
\end{equation}

The axiomatic Kijowski POVM therefore has exactly the structure obtained
from the minimal-current construction: the minimal positive current within
each directional sector, together with complete separation of the two
opposite momentum sectors.

\medskip
\noindent

\medskip
\noindent\textbf{Remark on boundary-constrained realizations.}
The result $K^{+-}=0$ applies to the unrestricted free particle, for which
the positive- and negative-momentum sectors are independent. If a physical
constraint relates the two momentum signs, this argument no longer applies,
and interference between them is not excluded. This occurs in the MS  construction  \cite{Marchewka2026}, where the boundary condition correlates the $+k$ and $-k$
components so that the problem is parametrized by a single independent
momentum variable $k>0$. On this reduced space, the normalized MS current
coincides with one directional Kijowski branch \cite{Marchewka2026}.
In the original momentum representation, however, each such reduced state
contains correlated $+k$ and $-k$ components, and their interference is
retained rather than eliminated. Moreover, the MS construction incorporates
a first-arrival condition through the absorbing boundary process, a
condition that is not contained in either the minimal current constructed
here or in the Kijowski POVM itself.

 \section{Discussion}

The initial motivation of this paper is to answer a simple question: what is
the minimal modification of the Schr\"odinger-current kernel required to
eliminate backflow? It is not obvious a priori that such a minimal
modification exists. We show, however, that the problem has a well-defined
solution, which we call the minimal current. Within each fixed-sign momentum
sector, the Schr\"odinger kernel
\begin{equation}
K_{\rm Sch}(p,p')=\frac{p+p'}{2m}
\end{equation}
is minimally modified to
\begin{equation}
K_{\min}(p,p')=\frac{\sqrt{pp'}}{m},
\end{equation}
while preserving its diagonal.

For opposite momentum signs, the same directional no-backflow requirement
gives
\begin{equation}
K_{\min}^{+-}=K_{\min}^{-+}=0.
\end{equation}
Thus a state containing both momentum signs does not define an additional
mixed arrival channel; its arrival current is completely determined by the
two directional sectors,
\begin{equation}
J_{\min}(t)=J_+(t)+J_-(t).
\end{equation}
This is the first main result of the paper, developed in Secs.~3--4.

Furthermore, the resulting minimal current defines exactly the Kijowski
time-of-arrival POVM, which was originally constructed axiomatically. This
is the second main result of the paper, presented in Sec.~5, and may be
viewed as providing a physical current-based motivation for both the
directional kernels and the separation of the two Kijowski branches.

The construction also clarifies an additional point. The absence of
interference between opposite momentum sectors is not a universal
requirement for every arrival-time POVM. It follows here because, for the
unrestricted free particle, the positive- and negative-momentum sectors are
independent and the directional no-backflow condition forces the
cross-sector kernel to vanish. If instead the two momentum signs are related
by a physical constraint, an arrival-time POVM with interference between
them is not excluded. This is precisely the situation in the MS
construction, where the boundary condition correlates the two momentum
components. In the normalized MS construction, the reduced current was
previously shown to coincide with one directional Kijowski branch
\cite{Marchewka2026}, while in the original momentum representation the
correlated $+k$ and $-k$ components retain their interference.

An important limitation of the minimal-current construction is that
positivity and normalization alone do not impose a first-arrival condition,
as discussed in Sec.~3.

The MS construction provides a complementary perspective. There the
first-arrival condition is built into the absorption process itself, and in
the normalized case the resulting directional current was shown to coincide
with one directional Kijowski branch \cite{Marchewka2026}. Thus the same
directional time-of-arrival density can arise either as the minimal positive
Schr\"odinger-current POVM or as the current associated with a genuine
first-arrival construction.

\end{document}